\documentclass[12pt]{article}
\usepackage{graphics}
\usepackage{graphicx}
\usepackage{epsfig}
\usepackage{latexsym}
\newcommand{\be}{\begin{equation}}
\newcommand{\ee}{\end{equation}}
\newcommand{\bea}{\begin{eqnarray}}
\newcommand{\eea}{\end{eqnarray}}

\begin{document}

%**************************************************
\baselineskip=.165in
%\baselineskip=.33in
%***************************************************

\begin{flushright} 
% hep-ph/0605218 \\
% LA-UR-06-3627 \\
\end{flushright} 

\begin{center}

\large{\bf 
Testing the Unitarity of the CKM Matrix with a Space-Based Neutron Decay Experiment}

\vspace{0.5in}

\normalsize
\bigskip 

Michael Martin Nieto${^a}$, William C. Feldman${^b}$, and David J. Lawrence${^c}$

\normalsize
\vskip 15pt

${^a}$Theoretical Division (MS-B285) and 
${^{b,c}}$Space and Atmospheric Sciences (MS-D466)\\ 
Los Alamos, New Mexico 87545, U.S.A.\footnote{Current addresses:  
$^b$Planetary Science Institute, 
1700 East Fort Lowell Road, Suite 106, 
Tucson, AZ 85719-2395. \\
$^c$Johns Hopkins Applied Physics Laboratory, 
11100 Johns Hopkins Road, 
MP3-E104, 
Laurel, MD 20723.}
\footnote{Email: $^a$mmn@lanl.gov, $^b$feldman@psi.edu, $^c$david.j.lawrence@jhuapl.edu} \\ 
\vspace{0.25in}

%***********************************************
% \today
%**************************************************

\vspace{0.5in}
\bigskip

\end{center}

%******************************************************************
% \baselineskip=.33in
\baselineskip=.165in
%******************************************************************

\begin{abstract}
If the Standard Model is correct, and fundamental fermions exist only in the three generations, then the CKM matrix should be unitary.  However, there remains a question over a deviation from unitarity from the value of the neutron lifetime.  We discuss a simple space-based experiment that, at an orbit height of 500 km above Earth, would measure the kinetic-energy, solid-angle, flux spectrum of gravitationally bound neutrons (kinetic energy $K<0.606$  eV at this altitude).   The difference between the energy spectrum of neutrons that come up from the Earth's atmosphere and that of the undecayed neutrons that return back down to the Earth would yield a measurement of the neutron lifetime.  This measurement would be free of the systematics of laboratory experiments. A package of mass $<25$ kg could provide a $10^{-3}$ precision in two years.  
\\

\end{abstract}

\begin{center}
\today
\end{center}

\newpage

%********************************** 1 Intro

\section{Introduction}

In the Standard Model of particle physics \cite{SM}, there are three generations of (doublet) quarks $[(u,d),~(c,s),~(t,b)]$ and three generations of electroweak leptons $[(e^-,\nu_e),~(\mu^-,\nu_\mu),~(\tau^-,\nu_\tau)]$.  Only quarks experience the strong interaction.  On the other hand both quarks and leptons experience the electroweak force.  The (strong interaction) mass eigenstates of the quarks are NOT the same as the weak eigenstates; there is mixing \cite{terry}.  

The mixing matrix is called the CKM (Cabibbo-Kobayashi-Maskawa) matrix.  By convention the charge $+2e/3$ $(u,c,t)$ quarks are taken to be unmixed, so the matrix acts on the charge 
$-e/3$ $(d,s,b)$ quarks.  The matrix can be parametrized as
\be
\left(\begin{array}{c} d'\\ s'\\ b' \end{array} \right) =
\left(\begin{array}
{ccc}
V_{ud} & V_{us} & V_{ub} \\
V_{cd} & V_{cs} & V_{cb} \\
V_{td} & V_{ts} & V_{tb} \end{array} \right) 
\left(\begin{array}{c} d\\ s\\ b \end{array} \right), \label{ckm}
\ee
where the mass and weak quark eigenstates are the unprimed and primed states.  (This is all like the case of the hyperfine states of the hydrogen atom, which become mixed and change their eigenvalues as an external magnetic field is turned on \cite{tan}.)  

If the Standard Model is the ``complete" theory (e.g., there is no new physics such as other fermion/neutrino sectors or extra dimensions), then $V$ must be unitary; i.e., the sum of the mod-squares of the elements in any row or column must be unity.  
In particular this applies to the first row of $V$, which is the row or column that presently can be most precisely tested,  
\be
U = |V_{ud}|^2 + |V_{us}|^2 + |V_{ub}|^2 \rightarrow 1.
\ee

But when this was looked at in detail, a small problem arose \cite{marciano}-\cite{pdg}.  
Given that the matrix elements $|V_{us}|$ and $|V_{ub}|$ were reasonably well determined from the decays of K and B mesons 
\cite{gilman}, by a few sigma unitarity still did not result.

But recent work on $|V_{us}|$ has yielded more precise experimental results \cite{newKL} and theoretical corrections \cite{marpdg07}, yielding 
\be
|V_{us}|= 0.2255 \pm 0.0019.   \label{sb}
\ee
(The effect of $|V_{ub}| = (0.00431 \pm 0.00030)$ is negligible.)   
Historically, the matrix element $|V_{ud}|$  was determined from the super-allowed beta decays of nuclei, with difficult strong-interaction corrections applied \cite{marsir}.  Over time 
these experiments have yielded improving results,  \cite{marpdg07,hardy,savard,hardytowner}, currently yielding
\be
|V_{ud}^N| = 0.9742 \pm 0.0003,   \label{VudN}
\ee
which in turn yields only about a one sigma deviation from unitarity \cite{marpdg07}

%********************** 2

\section{The neutron lifetime}
\label{nlife}

However, when the most elegant and direct measurement to obtain  $|V_{ud}|$, neutron beta-decay, is considered,   
\be
n \rightarrow p + e +\bar{\nu_e},
\ee
new problems arise.  This process simply involves the neutron to proton quark transition $(udd)$ to $(uud)$,  
\be
d \rightarrow u + e +\bar{\nu_e},
\ee
It proceeds via the interaction 
\bea
{\it I} &=& \frac{G_F}{\sqrt{2}} V_{ud}~\bar{e}(1-\gamma_5)\nu_e~
\bar{u}(1-\gamma_5)d, \\
 \frac{G_F}{\sqrt{2}} &=&  \frac{g^2}{8M^2_W}, ~~~~~~
 g \sin\theta_W = e,
\eea
where $G_F$ is the Fermi coupling constant, $M_W$ is the $W$-boson mass, $\theta_W$ is the Weinberg angle, and $e$ is the electric charge.

Neutron beta-decay experiments have much smaller theoretical uncertainties than nuclear beta decay experiments \cite{marciano,gilman,marsir}.  Therefore, the numerical value of the neutron lifetime $\tau_n$, is a more direct measure of the quark transition matrix element $V_{ud}$.  It also has great significance for the physical world that we live in \cite{schramm,bahcall}. First of all, it determines the reaction rates in the proton-proton chain and thereby the energy production in stars such as our Sun.  Even more fundamentally, the same physics determines the amount of primordial helium, which also affects the histories of later generations of stars. 

\begin{figure}[h!]
 \begin{center}
\noindent    
\psfig{figure=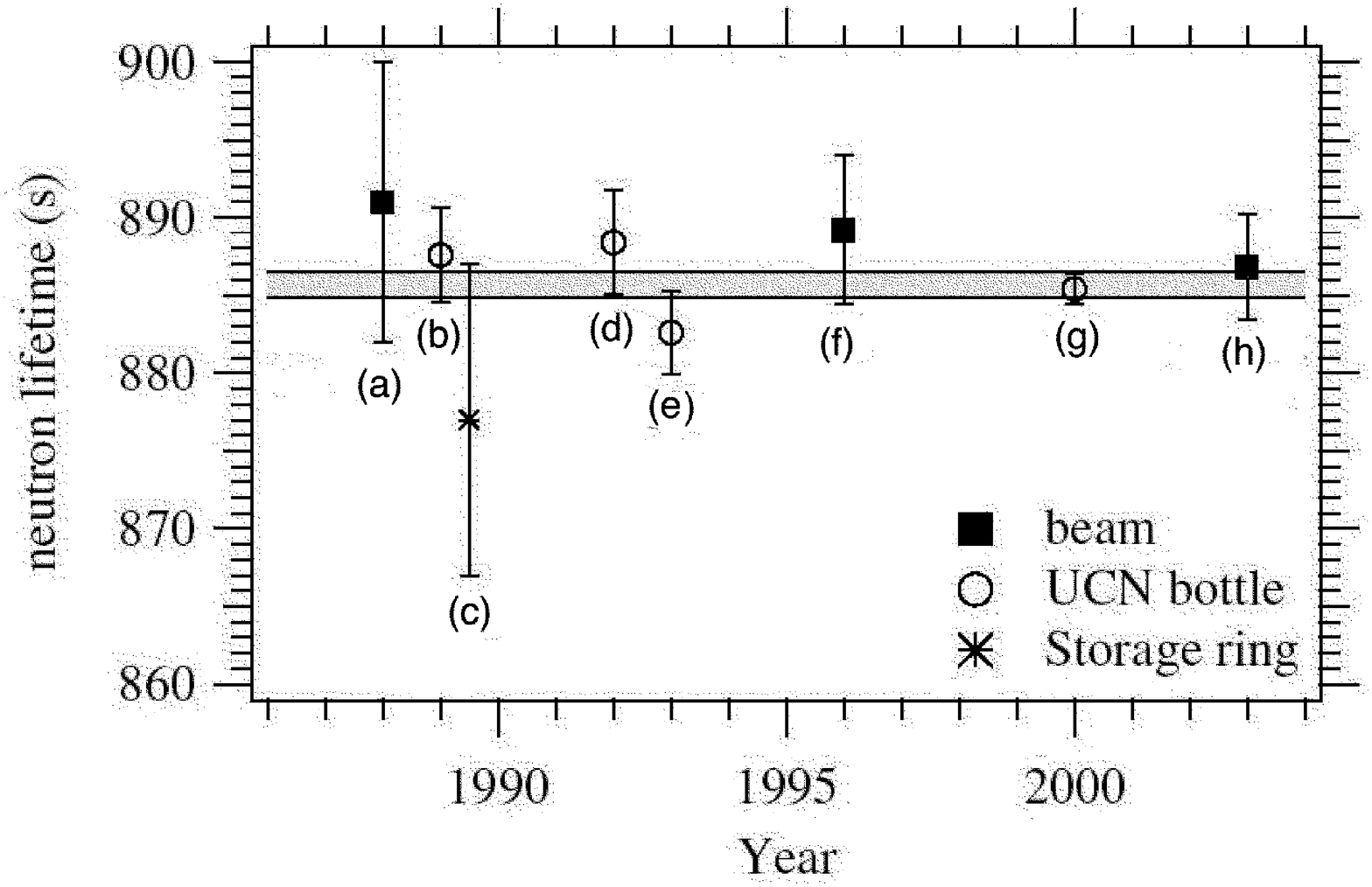,width=4in}%,height=90mm}
\end{center}
\caption{\small{Modified from Ref. \cite{h}, a comparison of recent neutron lifetime measurements, labeled (a) - (h), given in Refs. \cite{a} - \cite{h}.
The gray band is the 1-$\sigma$ deviation about the weighted result.}
\label{ndecayplot}}
\end{figure}

As recently as the late 1980's, the various measured values of $\tau_n$ deviated by as much as 7\% even though the individual accuracies were quoted in the 1\% range.  This was mainly due to the difficulty in determining the experimental systematics involved in the various experiments.  But since that time experiments have improved to yield accuracies as low as in the 0.1\% range \cite{a}-\cite{h}.  This yielded a 2004 world average of 
\cite{pdg}
\be
\tau_n = (885.7 \pm 0.8)~ \mathrm{s}.
\ee
Figure \ref{ndecayplot} shows these results.  

%******* Fig 1 

%******************

With this value, there arose a slight inconsistency with the Standard Model.\cite{gilman}.   Using the 2004 accepted number for $|V_{us}|= (0.2200 \pm 0.0026)$ \cite{gilman}, and the value for $|V_{ud}^n| = 0.9725 \pm 0.0013$ taken from neutron decay alone \cite{gilman},
the deviation of $U$ from unity was slightly over 2-$\sigma$, 
\be
U^n = 0.9942 \pm 0.0028,
\ee
with superscript ``$n$" meaning from neutron decay.

However, it must be remembered that in 2004 i) part of the error came from $|V_{us}|$ \cite{gilman,newKL}.  Further, ii) part of the problem was due to the fact that \cite{abeleprl}
\be
|V_{ud}|^{-2} = \tau_n\left(\frac{G_F^2m_e^5}{2\pi^3}\right)
         (1 +3\lambda^2)f^R(1+\Delta_R), 
\ee
where  the (model-dependent) phase space factor$f^R$, the radiative corrections $\Delta_R$, and especially $\lambda = G_A/G_V$  all had their own errors.  When these things were both were addressed, as partially discussed above \cite{marpdg07}, the problem with unitarity seemed to disappear.  

But then the most recent gravitational trap experiment \cite{2005} gave a surprising result in strong disagreement with the previous world average: 
\be
\tau_{n_{GT}} = (878.5 \pm 0.7_{stat} \pm 0.3_{syst}) ~\mathrm{s}.
\ee
This result disagrees with the current conditions for unitarity.  The situation obviously remains unsettled.  

Indeed, the Particle Data Group's evaluation \cite{pdg07} of experiment 
\cite{2005} is that, ``The most recent result, that of  \cite{2005}, is so far from other results that it makes no sense to include it in the average. It is up to workers in this field to resolve this issue. Until this major disagreement is understood our present average of $(885.7 \pm  0.8)$ s must be suspect." 

What we are emphasizing is that, since the neutron experiments are mainly neutron bottle or beam measurements, each with their own experimental systematics, the question arises whether one could tie this down better with a different type of experiment.\footnote{Note, also, that new experiments using ultra-cold neutrons in magnetic traps have been proposed \cite{bowman, TUM}.}
In particular, it could help determine more precisely if the proper number is close to the neutron lifetime that yields unitarity.

% **************************************** 3

\section{A space-based neutron lifetime experiment}
\label{nspace}

We now describe a space-based measurement of the neutron lifetime which would have none of the systematics of laboratory experiments.  Those that would exist are conceptually very different.  This would provide a new way of looking at the problem.

\begin{figure}[ht!]
 \begin{center}
\noindent    
\psfig{figure=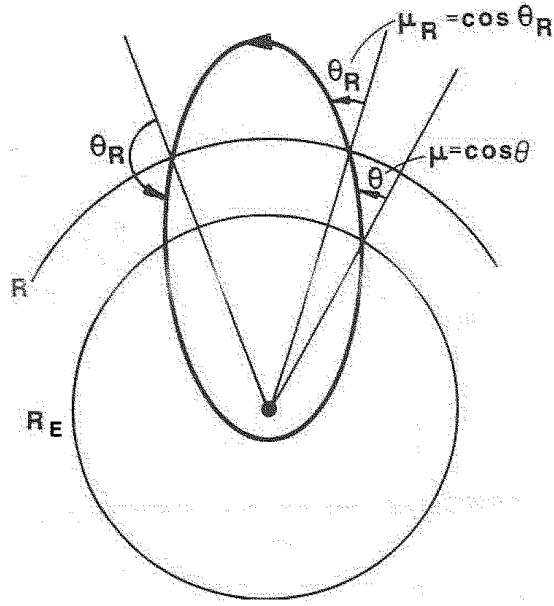,width=3in}%,height=90mm}
\end{center}
\caption{\small{Orbit of a gravitationally bound neutron about the Earth.
(Adapted from \cite{feldman2}.)  $R$ is the radius of the satellite's orbit about the Earth. $R_E$ is the radius to the top of the Earth's atmosphere. $\theta$ is the angle of of the neutron's elliptical orbit (with focal point at the center of the Earth) with respect to the normal at the Earth's surface.  $\theta_R$ is the similar angle with respect to the satellite's orbit.  
}
\label{norbit}}
\end{figure}

The fundamental idea of this experiment was described in Ref. \cite{feldman}.  (In Figure \ref{norbit} we show the geometry of the system.)
It depends on the fact that the neutrons created in Earth's atmosphere by cosmic rays, and leaving the atmosphere with kinetic energies, $K(R_E)$, below 0.65 eV, are gravitationally bound \cite{feldman2}.  Further, of these, some survive going upward to  $h = 500$ km altitude, or orbital altitude $r= R\equiv h + R_E$.  They now have $K(R)< 0.606$ eV, and are still gravitationally  bound.  After having passed this particular orbital altitude on the way up,  some of these neutrons will have decayed by the time they fall back down to the same altitude.  Therefore, if one has neutron detectors in a circular Earth orbit, the difference between the energy spectra of up-going and down-going neutrons with energies below 0.606 eV is a measure of the neutron lifetime.  

%****************** Fig 2

%******************

%****************** Fig 3

\begin{figure}[h!]
 \begin{center}
\noindent    
\psfig{figure=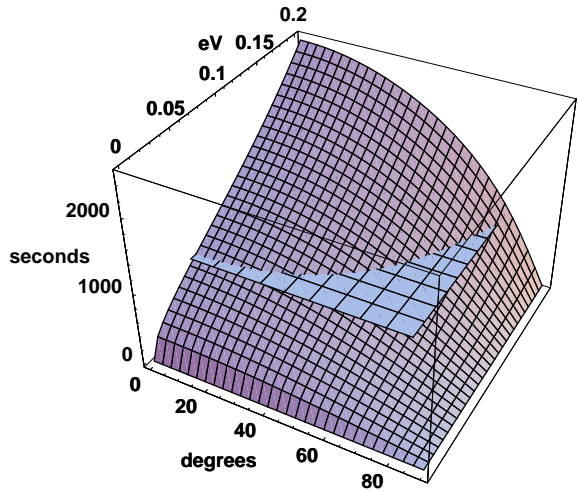,width=3in}%,height=90mm}
\end{center}
\caption{\small{ Round-trip travel time, $\Delta t_{RR}$, up from  $R$, the radius of the satellite's orbit about the Earth and back down again to $R$:  $\Delta t_{RR}$ is plotted in seconds ($z$ axis).  It is plotted as a function of angle, $\theta_R$, in degrees ($x$ axis) and kinetic energy, $K$, in eV ($y$-axis).  A slice shows the 2004 world average value of the neutron lifetime, $\tau_n = 885.7$ s.}
\label{ndecayT}}
\end{figure} 

%******************

Integrating Newton's equations, the time for a bound neutron of mass $m_n$, with kinetic energy $K(R)$ at the height of the satellite's orbit, 
\be
K(R) \le K_{max}(R) = -V(R) =  \frac{m_nGM_E}{R} = 0.65~ \mathrm{eV},
\ee
to travel from the radial distance of the satellite, $R$, to a general position, $r$, can be determined analytically.  In particular, the time needed to return back down to the same height, $R$, is found to be
\bea
\Delta t_{RR} &=& \frac{R \sqrt{m_n/2|V(R)|}}{[1 - K(R)/|V(R)|]^{3/2}}
\left[\frac{\pi}{2} + A +\sin^{-1}\left(\frac{B}{\sqrt{A^2+B^2}}\right)\right],\\
A&=& \left[4\left(\frac{K}{|V|}\right) 
\left(1 - \frac{K}{|V|}\right) \mu_R^2\right]^{1/2}, ~~~~~~
B = \left[ \frac{2K}{|V|} -1\right].
\eea
where $\theta_R= \cos^{-1} \mu_R$ is the angle with respect to the radial.  

The round-trip time for an up-going neutron, with a kinetic energy of a few tenths of an eV, is of the order of the neutron lifetime.  (See Fig. \ref{ndecayT}.)  Therefore, by measuring the overall upward energy spectrum and comparing it to the downward spectrum, an exponential decay lifetime can then be obtained. In Figure \ref{ndecayT} we show the round-trip time as a function of angle $\theta_R$ and kinetic energy $K$.  

In addition to its conceptual simplicity, the main advantage of this method is that 
it has none of the systematics that, by construction, plague previous measurements \cite{byrnebook}.  
Significant systematics include neutrons created in the spacecraft
itself, high-energy gamma-rays, detector efficiencies, and the
time-variations of the Earth's atmosphere, magnetic field, and incoming
cosmic ray flux.  With effort these effects can all be accounted for and
the expected counting rate can be predicted rather well as has been shown
from neutron measurements from the Moon \cite{lawrence06}.

\begin{figure}[h!]
 \begin{center}
\noindent    
\psfig{figure=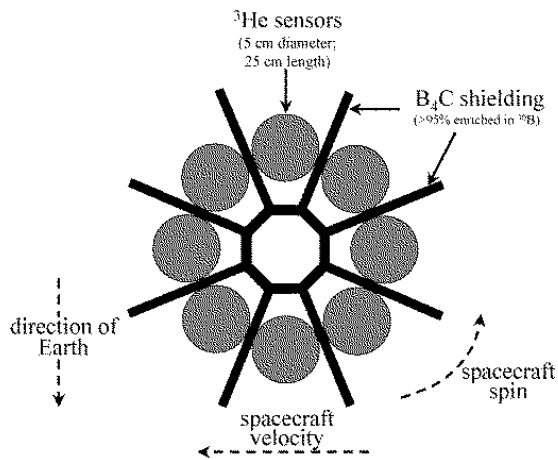,width=4in}%,height=90mm}
\end{center}
\vskip -30pt
\caption{\small{Diagram of sensor}
\label{ndecaysensor}}
\end{figure}

Space-based neutron sensor techniques
have been used to make measurements of hydrogen abundances on the Moon and
Mars \cite{feldman98,feldman02}.  The experiment here
consists of an array of eight of the same type of $^3$He gas proportional
counters (5 cm diameter $\times$ 20 cm long active volume) used for the Moon
\cite{feldman04}.\footnote{A gas pressure will be used that is sufficiently large to detect all of the gravitationally bound neutrons efficiently but not too large to contaminate the measurements with higher-energy neutrons. 
}
They are arranged in a cylindrical geometry with each sensor
collimated to a 45$^\circ$ azimuthal acceptance angle using eight planar
`fins' composed of sintered $^{10}$B-enriched boron carbide.
In Figure \ref{ndecaysensor} we show a diagram of this arrangement.

%****************** Fig 4

% ***********************************

Placing this sensor on a spin-stabilized micro-satellite having spin axis
perpendicular to the orbital plane with a rotation of rate of $\sim5-10$ rpm, then would allow a measurement of the full energy-angle, atmospheric-leakage, neutron flux function,  
using a Doppler filter technique \cite{doppler}.\footnote{In a nearly inertial polar orbit the solar panels would be placed on the front face of the detector as shown, constantly facing towards the Sun.}  In this technique, use is made of the fact that in low-Earth orbit satellites can travel faster than thermal neutrons.  For example, the orbital velocity at an altitude of 500 km is 7.6 km/s whereas the velocity of a 0.1 eV thermal neutron is 4.4 km/s.  This means a forward facing and a rearward facing sensor experience vastly different spectra, with intermediate spectra at orientations in between.  Combined with a rolling spacecraft, this allows the spectra to be well measured.  

Simulation of the counting rate when one of the sensors is facing upward (detecting downward-going gravitationally-bound neutrons) yields a counting rate  of about $2.2 \times 10^{-4}$ cm$^{-2}$s$^{-1}$, when the craft is poleward of $60^\circ$ magnetic 
latitude.\footnote{The thermal-energy leakage neutron flux is relatively constant and largest in these two polar cones \cite{highEn}.  A ninth cylinder might be placed in the center of the array in Figure \ref{ndecaysensor} to provide direct calibration of this high-energy neutron flux.}
Similar simulations have shown \cite{feldman} that to obtain a
measurement of the neutron lifetime to 
a precision, $p$, one only needs the number of downward counts to be
\bea
N(p) &=&   \frac{1}{2 p^2}        \\
N(10^{-3}) &=& 5 \times 10^5.   \label{countrate}
\eea

Using this, we find that in two years the proposed $^3$He array could obtain the $5 \times 10^5$ downward counts needed for the precision of $10^{-3}$ given in Eq. (\ref{countrate}).  This gives confidence in the viability of the experiment, which we emphasize would directly measure the up and down spectra independently of any simulations.  

This sensor is very easy to field, should be relatively
light weight (less than about 25 kg), and can be operated using very
simple, proven analog electronics.

% *************************** 4

\section{Conclusions}

Since this experiment uses mature and understood technology and software, it would be relatively inexpensive and straight forward to mount.  The experiment itself is an obvious candidate for multi-agency consideration.  Using a near polar orbit, the prime launch site for the mission would be Vandenberg Air Force Base in California.  The launch vehicle could even take more than one capsule aboard.  Since it would be a unique experimental input into the physics under consideration, this concept should be rigorously evaluated.

% *********************

\section*{Acknowledgements}
We thank Jim Friar and Terry Goldman 
for helpful discussions and comments on the theoretical 
corrections to the $V_{ud}$ matrix element. 
The support of the United
States Department of Energy under contract W-7405-ENG-36.is also acknowledged.

% ****************************

%***********************************************

\end{document}